\documentclass[12pt,preprint]{aastex}

\begin{document}

\title{Optical polarization observations of NGC 6231: evidence for a past SN fingerprint\altaffilmark{1}}

\author{Carlos Feinstein\altaffilmark{2}, Ruben Mart\'{\i}nez
%\altaffilmark{3}
, M. Marcela Vergne\altaffilmark{2}, Gustavo Baume \altaffilmark{2},  Rub\'en V\'azquez 
\altaffilmark{2}}

\altaffiltext{1}{Based on observations obtained at  Complejo Astron\'omico El
Leoncito (CASLEO), operated under agreement between the CONICET and the
National Universities
of La Plata, C\'ordoba, and San Juan, Argentina.}

%\altaffiltext{3}{On a fellowship from CONICET, Argentina.}
\altaffiltext{2}{Member of Carrera del Investigador Cient\'{\i}fico, CONICET,
Argentina.}

\affil{Facultad de Ciencias Astron\'omicas y Geof\'{\i}sicas, Paseo del Bosque, 1900 La Plata, Argentina} 

\begin{abstract} 
We present the first linear multicolor polarization observations for a sample
of 35 stars in the direction of the Galactic cluster NGC 6231. We have found a complex pattern 
in the angles of the polarimetric vectors.
Near the core of this cluster the structure shows a semi-circular pattern that we have interpreted 
as a re-orientation of the dust particles, showing the morphology of the magnetic field. 
We propose that a supernova event has occurred some time ago and produced a shock on the local ISM. 
We discuss in this paper independent confirmations of this event, both from the studies on the diffuse
interstellar absorptions \citep{cro01} and the results of the pre-main sequence stars (PMS) given 
by \cite{sung98}. We also show that a supernova is supported by the evolutionary status of the cluster.
\end{abstract}  
       
\section{Introduction}  

The polarimetric techniques are a very useful tool to find important information
\break ($P_{\lambda max}$,$\lambda_{max}$, magnetic field direction, etc) from the dust located
in front of a star (or a luminous object), or from dust located inside a stellar cluster.
Young open clusters are very good candidates to carry out polarimetric observations, because
 previous photometric and spectroscopic studies of 
these clusters give detailed information of the main sequence, so we can characterize the
physical parameters of the member stars (age, distance, memberships, etc),  
in order to study the extinction of the dust in the direction to the cluster and within the cluster.
%In this framework, we are doing systematic polarimetric observations of a large number of clusters in the galaxy. 
%Polarization of open clusters is very interesting because of the amount and quality of information that is obtained. The degree of polarization (and the angle of orientation) of the light depends on the  ISM toward the cluster, the ISM within the cluster and  the intrinsic properties of the individual stars of the cluster. In combination with photometry and spectroscopy a large amount of information of the dust, the stars and the cluster itself can be obtained. 
For example, from the study of the $P_{\lambda max}$ versus the $E_{B-V}$ of each star of a cluster it is 
possible to 
find the polarimetric components of the extinction produced by Galactic dust located between the Sun and 
the cluster (e.g. Tr 27, \citet{fein00}). 
By subtracting the effect of this dust on the line of sight from the data it is possible to study the 
component associated with the internal extinction of the cluster. In addition to 
this valuable information, stars with intrinsic polarimetric properties can be isolated. 
%From the fitting of the Serkowski law to the multicolor data properties of the dust connected with the size and composition of the dust are found (e.g. ...)
The angle of the polarization vector is an estimator of the direction 
of the Galactic magnetic field. By observing members and nonmembers stars (foreground and background) 
the direction of the magnetic field towards the cluster at several distances can be inferred.

In a typical open cluster the polarimetric data show, for ``normal'' stars (that is, not considering stars 
with intrinsic polarization), a distribution of angles that can be described   
 by a mean fixed value which characterizes the cluster. The polarization angles of the cluster members show some scatter 
($\sim 10 ^{\circ}$ or less) 
over this mean angle, mainly because of density inhomogeneities in the dust along the line of 
sight to the cluster, inhomogeneities of the dust inside the cluster, to the intrinsic 
polarimetric properties of the stars and also due to more complex magnetic field structures.
For example, in Tr 27 the polarimetric data show a mean angle of 
$\theta\sim 30^{\circ}$ for the cluster with a FWHM of $\sim 10^{\circ}$ \citep{fein00}, in Ara OB1 
the polarimetric observations give a FWHM of $\sim 9^{\circ}.9$ for NGC 6167, $\sim 13^{\circ}.1$ for NGC 6193 
and $\sim 9^{\circ}.5$ for NGC 6204 \citep{wald99}. In the case of Stock 16 \citep{fein03} 
the FWHM is  6$^{\circ}$.7, which seems to be slightly smaller than the other clusters.

%Due to the valuable information that is obtained with the polarimetric studies of open clusters, we are carrying out a program of systematic observations of this kind of objects.In this paper we will to show a quite unexpected result obtained for the cluster NGC 6231, whose polarimetric data appear very different from those found in other open clusters. 

NGC 6231 (l= 345$^\circ$.5 , b=1$^\circ$.2) is known in the literature to be one of
 the youngest open clusters in the Galaxy and one of the brightest in the Southern
 Hemisphere. The cluster is embedded in an extended system of O and early-B stars, 
namely the very young association Sco OB1, that contains the ring-shaped HII region
IC4678, centered on NGC 6231. Usually, NGC 6231 is considered as the nucleus
 of Sco OB1 \citep{per91}. Its age  determinations range between $3 \times  10^{6}$ 
years \citep{mer86} and $ 6.9\times 10^6 $ years \citep{per91}. 
It is known as one of the richest open clusters, containing WR stars,
 $\beta$ Cepheid variables and  more than 100 O and B stars \citep{sho83}. According to 
photometric CCD observations of \citet{sung98} and of 
 \citet{bau99} there is  evidence of a population of pre-main sequence stars (PMS).
%The distances  given to NGC 6231 can be grouped into two intervals. The first one includes  the smaller distance values (about 1300 pc) \citep{mer81, sho83, van84, balo95}. The second one includes the higher distances between 1800 pc and 2100 pc \citep{bok66, seg68, fein68, sch69, per91}. 
The distance obtained by deep CCD photometry by \cite{bau99} locates the cluster NGC 6231 at $1990 \pm 200$ 
pc from the Sun.

%On the basis of spectroscopic studies \citet{gar01} have found in NGC 6231 a hight percentage in binary or multiple star systems (global binary frequency of $82 \%$ for stars earlier than B1.5V and $79 \%$ for the O-type stars).

The average visual extinction toward the cluster is about $A_{V} \sim 1.4 $ \citep{balo95,bau99}.
According to \citet{rab97} in the northern part of the cluster the absorption is constant, 
but it increases from the central part towards the South. A
similar trend, but less pronounced, is observed from West to East in the 
southern part. Part of the absorption suffered by the light from the cluster stars is due to
matter within the distance interval 100-1300 pc from the Sun 
\citep{per91}, 
but the field shows a relatively good transparency between about 400 and 1000 pc at a 
constant value averaging of $A_{V}=0.75$ mag (Raboud et al. 1997).
%On the other hand, the differential reddening and the large scatter of color excess (see Baume et al. 1999) may be due to several factors such as anomalies of the intracluster material and the presence of many binaries and variables stars. 

%Near the center of NGC 6231 there is a hole of reddening material, as a result 
%of the action stellar winds from massive stars about the material 
%surrounding PMS stars. This hypothesis is supported from \citet{bok66}, 
%they show an elliptical ring centered on NGC 6231.

\section{Observations}

The bulk of the observations were obtained on two observing  runs. The first one
was performed during 1995 (May 30 to June 3). The second one was on 
3 nights (August 17-20) in 2001. Two more runs on May 2002 and on August 2002 were used for re-observing 
and checking some stars.
All runs were carried out using the five-channel photopolarimeter of the Torino Astronomical Observatory
 \citep{scal89} attached to the 2.15 m telescope at the Complejo Astron\'omico El Leoncito
(San Juan, Argentina). Polarimetric measurements have been made for 36 stars to the cluster NGC 6231. 
Most of them were observed through the Johnson broad band UBVRI 
filters (${\lambda}_{Ueff}=0.360{\mu}m, {\lambda}_{Beff}=0.440{\mu}m, {\lambda}_{Veff}=0.530{\mu}m, 
{\lambda}_{Reff}=0.690{\mu}m, {\lambda}_{Ieff}=0.830{\mu}m$). 

Our results are listed in Table 1 which shows, in self explanatory format,
the stellar identification as given by
\citet{seg68}, the average of the percentage of polarization ($P_{\lambda}$)
and the position angle of the electric vector ($\theta_{\lambda}$) observed through
 each filter, with their respective mean errors.

The Torino polarimeter collects data for all the filters simultaneously (UVBRI), 
so all the wavebands have the same exposure times, and the S/N varies between filters.
Therefore, the values from different filters may be of different quality. Observations whose 
values are not over the 3$\sigma$ error level were discarded, and they are not reported in 
Table 1 neither used in this work. 

%Due to the nature of the results of this paper we decided to give a detailed explanation of the calibration process and to check the results of the observations between runs and with other published data.

Several standard stars for null polarization and for the zero point of the polarization 
position angle were measured for calibration purposes. At least three  
standard of angle stars were measured every night at different times. 
Errors were handled as described by \cite{mar92}.

In order to verify the lack of systematic differences between both main runs we have re-observed 
a common group of stars in the second run. 
%Fig. 1 shows the comparison of the resulting $P_{V}$ and polarization angle $\theta_{V}$ for this group of stars. Notice that the straight lines in this figure (with a 45$^{\circ}$ slopes) are drawn as a reference. As can be seen, in this figure no systematic  difference between the two runs are apparent.

Three other bright stars have been observed previously by \citet{serk75}: these are
HD 152234 (star \# 290), HD 152235  and HD 152248. Although
in that paper no observation errors were given, their  data 
(polarization and angle) are very similar to those reported in the present paper (see Table 2).

To look for nearby objects, the observed stars were checked with the Hipparcos/Tycho Catalog Data. 
Stars \#80,110,112,220,290 and HD 152248 were included in these catalogs but are 
so far away from the Sun that no useful parallax measures could be obtained.

\section{Results}    

Figure 1 shows the sky projection of the V-band polarization for the stars in NGC 6231. 
The dot-dashed line indicates the Galactic parallel $b=1^{\circ}.1$.
The first striking 
feature of this figure is the abnormal distribution of the polarization angles. Commonly
 in a typical open cluster the polarization vectors have a similar mean 
orientation of the polarization angle, 
and the individual stars show little scatter over this fixed direction. 
Previous studies for 5 clusters indicate that this scatter is equal or less than 10 
degrees of FWHM \citep{wald99,fein00,fein03}. As seen in Fig. 1, this is not the case in
NGC 6231.

%This is clearly not the case, it seems that there is not any orientation of preference for the polarization vectors. 

It is possible to classify the stars of NGC 6231 in three groups according to their location and the angle of the polarization 
vectors as shown 
in Fig. 1. Stars in  the northern part  of the image in Fig. 1 form the first group, 
 which has vectors orientated in the direction of the Galactic disk, see for example stars \#1,6,16,184 and 189.
To the South three more stars \#34,102 and 266 also have vectors orientated with the same direction.
Stars \#16,102, 184 and 189 are considered as nonmembers of the cluster \citep{bau99} and this fact can explain the
alignment of the polarization vectors with the Galactic plane. 
We believe that all this group probably consists of nonmembers or 
less reddened members which are closer to the Sun (``front-side stars'') and we are seing 
them without the effect of the intra-cluster dust. 

It is very noticeable that most of the polarization vectors of the stars in the inner core of the cluster seem to 
form part of a semi-circular structure and none of them appears aligned with the direction 
of the Galactic disk. We consider these stars as the second group.

The third group consist of stars with their polarimetric vectors at right angles 
to the direction of the 
Galactic disk. These are stars \#70,73,80,105,110,112,248,253,254 to the South. 
%However, there is no a sharp frontier between groups two and three. For example, stars \# 224 and CPD -41$^{\circ}$ 7733 are part of the semi-circular structure and have their polarimetric vectors with  the same direction as stars \#232,238,259,261 which are in the middle of these two groups.
 
The second group is a very unusual finding. In the next section we will discuss some 
probable explanations based on our polarimetric observations and on data from other authors obtained 
with different astronomical techniques (photometry, spectroscopy, etc).

\section{Analysis and discussion}

\subsection{Serkowski law}
%ley de Serk
To analyze the data, the polarimetric observations were fitted using  the 
Serkowski's law of interstellar polarization  \citep{serk75}. This is:

$P_{\lambda}/P_{\lambda max}=e^{-Kln^2(\lambda_{max}/\lambda)} \  \  \ \ [1]$

If the polarization is produced by aligned interstellar dust particles, the 
observed data (in terms of wavelength, UBVRI) will follow [1], where each star 
is characterized by  $P_{\lambda max}$ and  $\lambda_{max}$.

Adopting $K = 1.66  \lambda_{max}+ 0.01$  \citep{whittet92}, we have fitted our
 observations and computed the $\sigma_{1}$ parameter (the unit weight error of
the fit) in order to quantify the departure of our
data from the ``theoretical curve'' of  Serkowski's law. A $\sigma_{1}$ value larger
 than 1.5   is considered as an indication of the presence of
a component of intrinsic stellar polarization. Another criterion of intrinsic 
stellar polarization is to compute the dispersion of the position angle    
for each star normalized by the average of the position angle errors 
($\bar{\epsilon}$, Marraco et al. 1993).

The  $\lambda_{max}$ values can also be used to test the origin of the polarization. 
In fact, since the average value  of $\lambda_{max}$
for the interstellar medium is 0.545 $\mu m$ \citep{serk75}, objects showing  $\lambda_{max}$
rather lower than this value are also candidates for having an intrinsic 
component of polarization \citep{orsa98}.
The values we have obtained for $P_{\lambda_{max}}$, the $\sigma_{1}$ parameter, 
$\lambda_{max}$, and $\bar{\epsilon}$ together with the identification of
stars are listed in Table 3.       

Figure 2 shows the observed $P_{\lambda}$ and $\theta_{\lambda}$ vs ${\lambda}$ 
of those stars which are 
candidates for having an intrinsic component of polarization (stars \#70,73 and 254). For comparison purposes, 
the best Serkowski's law fit has been plotted as a continuous line.
Star \#220 also shows an intrinsic component of polarization on Table 2, but it is already known as polarimetric
variable. First \citet{lun} and then \citet{st87} have made an extensive polarimetric study of 
this double-line binary (WC7 + O5-8) system.

\subsection{Polarization efficiency}

Fig. 3 shows the plot of  $P_{\lambda max}$ vs.  $E_{B-V}$ \citep{sung98,rab97,bau99} 
for each of the three groups. 
This plot is very useful for study the polarization efficiency, defined as the ratio 
$P_{\lambda max}/E_{B-V}$
which indicates how much polarization is obtained for a certain amount of extinction.

To understand the behavior of the polarimetry and the extinction 
 a detailed discussion of the distribution of the dust on the line 
of sight to the cluster is necessary. 
Neckel \& Klare (1980) have studied the extinction values of the ISM and distances in our 
Galaxy using more that 11.000 stars. Their Figure 6j(211) (which is the plot in 
the direction of NGC 6231) shows two steeps
in the absorption increase: the first one is very near to the Sun and associated with the Lupus cloud 
\citep{cro00} at 
150 pcs ($A_{V}=0.8$) and the second step is about 700 to 1000 pc and is also detected in the diffuse 
interstellar lines \citep{cro01}. This second step of absorption is responsible for $A_{V}\sim0.6$. \citet{bau99} have reconstructed Neckel \& Klare's plot with new CCD data, and found a more complex
structure obtaining a similar result for the distribution of dust between the Sun and the cluster. 
Nevertheless, Figure 7 of \citet{bau99} shows that NGC 6231 has also intracluster dust.

Most of the studies show that the cluster has  differential reddening. For example,
\citet{sung98} have made a bidimensional map of the distribution of $E_{B-V}$ over the NGC 6231 cluster 
(see Fig. 5 of their paper). They show that the core of the cluster is less reddened  
and seems to lie in a hole in the reddening material ($E_{B-V} \sim 0.45$). To the South 
they find that the extinction increases to $E_{B-V} \sim 0.65$. The differential reddening of
the cluster is extended discussed in \citet{rab97} and \citet{bau99}.

%############################# Corregir ###############################################################
%In Fig. 3 we have used different symbols for each of the three groups in which we have classified the
%stars of NGC 6231 according to their polarization vectors. The first and the third group are located 
%in the plot with larger values of polarizations and
%compatible with a  $P_{\lambda max}/E_{B-V} \sim 5$, meanwhile the second group is  
%in a place of low polarization efficiency. We have interpreted  this plot in terms of 
%different layers of interstellar polarization as seen in the path to the cluster \citep{nec80,bau99}
%and according the map of $E_{B-V}$ of \cite{sung98}. 
%The first group has polarizations compatible with the direction of the galactic disk and
%they are most probably produced by the interstellar dust in the line of sight,  which is 
%aligned in the direction 
%of the Milky Way disk. This dust is very unlikely to be related to NGC 6231 (or Sco OB1) because
%it must be located in an undisturbed place. At the North of the cluster, this layer seems 
%to dominate the composition of the different orientation of  
%polarization. So, the result is polarization vectors in the direction of the galactic disk but with
%some degree of depolarization due to the others components. 
%############################ Corregido ###############################################
Fig. 3 is the plot of  $P_{\lambda max}$ vs $E_{B-V}$ for each group of stars. 
The upper panel are stars of the first group, the middle panel is for second group of stars and 
the bottom panel are data of the third group of stars.
The solid line in the first panel of Fig. 3 is the empirical upper limit relation 
for the polarization efficiency by the interstellar polarization $P_{\lambda max} = R\ A_{v} = 9 E_{B-V}$
(for normal dust, $R=3.2$) given by Serkowski et al. (1975). 
The dashed line ($P_{\lambda max}/ E_{B-V} = 5 $) represents the observed
normal efficiency of the polarizing properties of the dust (Serkowski et al., 1975).

Nevertheless, the $E_{B-V}$ data were collected from different sources and probably some of these values
have large errors, some qualitative properties emerge from the figure.
It is noticiable that all the groups have a polarizing efficiency 
lower than the  $P_{\lambda max}/E_{B-V} \sim 5$ expected. Because for an extinction of $E_{B-V}=0.43$ \citep{bau99} and the
average polarization efficiency, the typical value should be around $P_{\lambda max} \sim 2\%$. This value is  much higher  than the one observed in the cluster (see Table 3).
Therefore, in all the groups we have an important depolarization and this efect is notorious for the
stars of the semi-circular structure (group 2) that have a much lower polarization 
in comparison with the first and third groups (see Fig. 3 middle panel). There are not stars with polarization
greater than 0.95\% in this group. 

%We interpreted these behavior due to a strong  depolarization because there is a cancellation  in the resulting polarization vector composition due to crossed fields along the line of sigh to this group. This cancellation happened to the data of all the stars in the cluster, but to the North the Lupus cloud dominates the composition result, to the South a cloud near 1 kpc dominates the composition and at the center of the cluster both Lupus and the other cloud cancel each other.

%############################### saco porque repite ##################################################
%To the South other dust layer dominates the polarization vector composition, the resulting angle of the polarization vector is not in the galactic disk direction. 

%\cite{sung98} and \cite{rab97} have shown that the extinction is variable over the cluster and increase to the South. This result is compatible with the idea that the observed polarimetry is the composition of components of at least two layers of dust. Therefore, as the reddening increases to the South a different polarimetric component dominates the observed angle in this direction.

%###############################

%Nevertheless, the observed depolarization, the typical trend of the grow  of $P_{\lambda max}$ when the $E_{B-V}$ increases, is still observed ($P_{\lambda max}/E_{B-V} \sim 5$).

%$P_{\lambda max}/E_{B-V} \sim 5$ for the average interstellar dust is still observed (see the dashed lines).

There is a question that remains to be answered. 
Why we can see this semicircular structure behind two layers of dust which orientate the polarization 
vectors in different directions. It is pretty obvious that just having only one dust layer on the line of sight  
would compose and change the direction of the polarization vectors, washing out this semicircular 
structure. But on the other hand, the observed semicircular structure seems quite perfect.
As we have discussed in the previous paragraph, in the particular case of NGC 6231 there are two layer of dust, a 
first one at Lupus (hereafter the first component of extinction) and a second one in front of the 
cluster locate at 1 kpc, \citep{rab97,bau99,nec80}, hereafter the second component.
It is very important to notice that the two components are orthogonal and one of them 
can compensate or diminish the other. 
The first component is dominated by dust in the Galactic 
plane, which has an angle of polarization of $\sim 45$ degrees, the second component seems to be 
at right angles to  the  Galactic plane at $\sim$ 135 degrees. 
%In the Stockes' parameters space where   the polarization is characterized with the parameters $Q$ and $U$ (where $Q=P\ cos(2\theta)$ and $U=P\ sin(2\theta)$), these directions are basically  opposite to each other. The first component is at $2\theta=90$ degrees and the second at $2\theta=270$ degrees (due to the $2\theta$ in the expressions of $Q$ and $U$).
Fig. 3 shows a huge depolarization for the second group of stars.
We think that in the northern part
the first component dominates the observed polarization, while in the South the second component is the strong one. 
But in the core, where the semicircular structure lays, both components (first and second) 
cancel each other, therefore making a polarization ``window'' where we can observe inside the cluster.
This interpretation also explains the depolarization of the stars locate in the semi-circular structure
(second group). 

We tested this interpretation of the extinction toward NGC 6231, by analyzing the polarization of 
the nonmembers stars. For example, in the first group, \#102 is considered a F0III 
\citep{per91} star 
so it must be located at about 379 pc from the Sun and star \# 189 is a B9-8V \citep{rab97} being located 
at 1 kpc. Both stars are
behind the Lupus cloud and have  polarization angles similar to the direction of the Galactic disk.
Therefore, we can deduce that the Lupus cloud is responsible for the  polarization of the light in the 
direction of the Galactic disk. This cloud is reddening the stars behind by $E_{B-V}=0.25$ in average \citep{nec80}.  
For larger distances \citet{rab97} have found that the interstellar medium shows a relatively good 
transparency between 400 and 1000 pc. 

Star \# 105 is also a nonmember star but its polarization (and location in 
the cluster) is related with the third group of stars.  Assuming that this star is a main sequence object
reddened by normal dust it is possible to found an unique solution from the color-color diagram that gives 
an $E_{B-V}=0.52$. This value is larger than the estimate for the Lupus cloud and shows
that this object is located behind another layer of dust which is polarizing the stellar light at
 right angles to the direction of the Galactic disk. Therefore, this star is probably located behind the second cloud
at a distance of more 1 kpc from the Sun.

%In this plot the solid line is $P_{\lambda max}/E_{B-V} \sim 9$, which is known to be the maximum efficiency for the ISM to polarized, meanwhile the dashed line is $P_{\lambda max}/E_{B-V} \sim 5$ which is the normal efficiency for the dust in our galaxy. We plot these two lines from the origin of coordinates to use them as a reference. Is easily noticeable from the plot that the observed  polarization is to low for the extinction. One possible explanation of this behavior could be that we are observing through several layers of dust each of them  having the particles in different orientations. Each of these layers depolarized the light, so the total effect would be  extinction but not very much polarization. In this particular case it is very noticeable the lack of correlation between $P_{\lambda max}$ vs. $E_{B-V}$, from an statistical test we obtain a $\chi^{2}= 152.438 $ which means a 99\% probability that the polarization and extinction are not correlated. 

%###################### Hay anormalidades en la ley de Serkowski (ojo deberia haberlas) #####

\subsection{ The semicircular polarization pattern } 

As despicted in Fig. 1, the polarization vectors of the stars belonging to our second group (at the core 
of the cluster) displayed a  semicircular pattern. 
This is very unlikely to be the result of a distribution of polarization angles at random, because the pattern 
is well organized.
We think we are in the presence of an energetic phenomenon that has changed the magnetic field orientation,
resulting in a realignment of the dust grains, which manifests itself in a rotation of the polarization vectors.
We propose that a supernova event has occurred some time ago and produced a 
shock over the ISM. It has pushed the gas and dust (and the associated magnetic field), causing the dust 
grains to re-oriented to the new direction of the magnetic field. 
We will discuss in the next paragraphs the relevant questions associated with the proposed
supernova event. 

Where did this phenomenon occur ?.  Tracing  lines perpendicular to the most
conspicuous polarimetric vectors it is possible to obtain an approximate  center at the crossing point where 
these lines overlap each other. This center seems to be 
at $\alpha =16^{h}\ 54^{m}\ 22^{s}$, $\delta = -41^{\circ} \ 46^{'}\ 30^{"}  $, J2000.0 (Fig. 4).
%This is not a good method to make an estimation because  we do not have the whole circle, we have only about a half of it ($\sim$ 220 deg), therefore our estimation is suffering from strong systematic errors. 
This semi-circle is very regular and the vectors converge clearly to a center. 
%nevertheless that inhomogeneities in density can make a shock or a ionization front to travel at different speeds at different directions making a confused pattern (e.g. M16 Pillars, etc). 
We think this is a solid argument about the existence of center of this pattern, which means that the
energetic phenomenon was produced at a certain fixed location in the cluster. This supports  the idea 
that a supernova exploded in the near past of the cluster, 
located at the center of the crossing lines as shown in Fig. 4.

Could  a supernova event have happened in NGC 6231 ?
The earliest star on the main sequence, not considering binaries, is at present of spectral type O9 
(\#292, HD 326329), 
%id nuestro ?, no tiene no la observamos 
and there is evidence of more massive stars that had evolve out of the main sequence \citep{bau99}. Therefore, it is very probable that a star more massive than HD 326329 had
evolved to a supernova stage in the past. Also, \cite{bau99} have computed the initial mass function of 
NGC 6231 (hereafter IMF) with binary correction and they have obtained a slope of $x=1.14\pm0.10$, for 
$x=log(dN/\delta M)/log(M)$. This slope of the distribution allows the existence of a former star of about
  70 $M_{o}$ (an O3-O4) without any change in the parameters of the observed IMF. 
This means that one or few (because we are dealing with small number 
statistics) very massive stars could have lived in the past with no contradiction with the actual IMF.
So, the evolutionary status of the cluster population in NGC 6231 indicates that some massive star 
have evolved and probably became supernovae.

%\citep{sung98} have photometric data of 852 stars brighter than $V \leq 16$mag, \citep{rab97} has determine memberships and found that more 250 stars showing the cluster is very rich and have a large corona. So it is very probably that more early stars than the O9 have existed and have evolved to a supernova stage, in the near past. 

%poner calculo detallado de la funcion de masa. 

%poner preubas de supernovas.

%Sco 1, the large OB asociation where ngc 6231 is in his center, is surrounded by bright filaments of  H$\alpha$ emission.
%Poner lo de los rayos gamma.

%When did this phenomena happen ?. This is the most difficult question to answer. From polarimetry alone it is  not possible to date the supernova explosion. From the evolutionary arguments, it would depend on the spectral type of the progenitor, which is unknown. 
   
% (completar poniendo el tiempo de relajamiento del dust por campo magetico).

%################################Lineas interestelares########################################

Is there  other evidence on the ISM of this supernova event ?
Crawford and collaborators (e.g. \cite{cro89a}, \cite{cro89b,cro92,cro01})
 have made a very detailed study of the interstellar lines (NaI, CaII, KI) towards Sco OB1. 
Although these studies are related to the whole Sco OB1 association, these series of papers   
include in the analysis some of the brightest stars of NGC 6231. They have found several components 
and classified them into three major groups \citep{cro01}.  The first one consists of several components 
with radial velocities in the range $-20 \leq v_{helio} \leq 6\ km\ s^{-1}$ which are consistent with 
the galactic rotation 
along the 1900 pc to Sco OB1. All the clouds in this component are presumably foreground, 
and  unrelated with Sco OB1. For example components in the velocity range 
$-6 \leq v_{helio} \leq 6\ km\ s^{-1}$ are
probably associated with the Lupus molecular cloud complex at a distance of 150 pc.

The second group of interstellar lines found by Crawford (2001, and references therein) is the ``shell component'' 
which has velocities more negative than -20 $km\ s^{-1}$ (-22 $km\ s^{-1}$ is the 
velocity expected for the galactic rotation model at a distance of 1900 pc) and these clouds 
of interstellar matter seem to be related with the shells surrounding the Sco OB1 association.  
In a previous work \cite{cro89a} found a very low NaI/CaII relation in this component.  
This finding was explained by the removal of the Ca atoms from the grain surfaces by shock velocities
of $\geq 20\ km\ s^{-1}$. 
New data  \citep{cro01} indicate that the observed shell component has an upper limit for the temperature of the
order of $T=450^{\circ}K$ (as observed in the spectrum of HD 152235, a member of NGC 6231), which is low for a post-shock 
temperature ($T\geq 10^{4}$). But as \cite{cro01} noted, the cooling time of a shock for
the regular interestellar medium ($v_{s}\ = 20\ km\ s^{-1}, n_{H}\ = 10\ cm^{-3}$) gives $t_{cool}=6300\ yr$,
meanwhile the depletion time scale for the Ca to become part of the interestellar grains is
$t_{dep}\ =\ 1.6\times 10^{7}\ yr$. Therefore, shocks in the near past can account for the low NaI/CaII 
observed in the interstellar lines.

The third component of interestellar line has velocities more positive than those predicted 
by the galactic rotation models. 
In NGC 6231, HD 152249 exhibits an interstellar absorption component at a velocity of +19 $km\ s^{-1}$. Note that
HD 152249 is located in the core of the NGC 6231, and 1' to the South of the semicircular polarization pattern.
This component has a low NaI/CaII ratios 
($\leq 0.4$), which (as described above) is an indication of shocks. This component
could be located near the sun in the ``Upper-Centarus Lupus'' 
%(de Geus, 1992)
 or it could be an expanding interestellar structure related to NGC 6231.

Both second and third interstellar components exhibit the low NaI/CaII ratio and in our opinion 
it is a strong support for shocks in the region and gives clues over a turbulent scenario 
in the near past of NGC 6231. 
These interstellar line components are probably located in the cluster, as discussed by
Crawford in his series of papers, and were produced by shocks in the region. 
This point of view of Crawford is consistent with the interpretation of our polarimetric observations.

%########################## lugar y tiempo del evento ########################

\cite{sung98}, studying the Pre Main Sequence (PMS) stars in NGC 6231, have found a 
discontinuity in the number of stars near the MS band with ages between 30 Myr and 12 Myr.
 They have suggested that probably some PMS are not $H\alpha$ emitters and they are 
not detected by the typical $R - H\alpha$ vs. V-I diagram.
They explained the lack of $H\alpha$ emission suggesting that all the material 
surrounding the PMS stars,
material that would normally produce strong H$\alpha$ emission, could be swept away by stellar 
winds from massive stars, Wolf-Rayet stars or past supernova explosions.
% Acording to \cite{sung98} invesigation, this scenario may explain the hole seen in the reddening material near the center of NGC 6231 and apparently near the star \#290 (HD 152234) and HD 152233. 
Their result is also compatible and a strong support to our 
polarimetric findings.

%#########################  X-ray and gamma ###################################

Other interesting result showing some clues to support the idea of energetic events on the cluster 
NGC 6231 came from \citet{man96}. To explain the COS B diffuse emission of gamma-rays 
observations, \citet{man96} propose that 
strong winds in the cluster interact with the interstellar medium
creating the suitable condition for cosmic ray enhancement.

%\citet{cor99} have analyzed the Rosat X-rays observations. The region of NGC 6231  was observed with both HRI and PSCS cameras. The HRI data is useful for resolving the  stars in the cluster. \citet{cor99} found 35 sources in the HRI images and most of them are associate with the massive stars of NGC 6231. We have search for an image of the possible bubble in both HRI and PSCS camera \citep{cor03}, and we have not found any trace. But we think that better observations (having larger S/N) with the new X-rays instrumentation are needed.

\section{Other possible explanations of the semicircular polarization pattern}

We have proposed a supernova event to explain our observations, but we have analyzed other possible
interpretations. For example, a loop in the magnetic field would lead to similar observations,
however several papers \citep{nec80, bau99} show very clearly that the dust along the line of sight is distributed 
in layers
and there is no dust between these layers. Thus, there is not a continuum of dust to be aligned by
the magnetic field. Therefore, we have discarded this interpretation of the semicircular polarization pattern.

Other explanation for the observed polarization pattern might not need a supernova. Same observed bubbles
in the Milky Way are supposed to be made by the strong winds of Wolf-Rayet stars over a long
period of time. In some of these interstellar bubbles, like WR 101 \citep{cap02}, the Wolf-Rayet star
appears to be locate in an eccentric position close to the densest part of the nebula. This phenomenon could 
be explained by an inhomogeneous  structure of the ISM near the Wolf-Rayet star or from the motion of the star inside the shell \citep{cap02}. This would be
the case if the Wolf-Rayet star \#220 (HD 152270) is responsible for the bubble in NGC 6231. 
The location of this star is eccentric to the semicircular pattern.

%Other interpretation, arises from the change in reddening the cluster have from North to South, so this semicircular structure could explained because we would be seen the composition of the different directions of the polarimetric vectors from North to South and from West to East.  This explanation has the problem that the observations of the stars involved in the semi-circle structure, must have a higher $\sigma_{1}$ and $\bar{\epsilon}$ parameters in Table 3, than the others stars. These parameters   are very sensible to a vector composition because the result would suffer from a departure from a pure Serlowski's  law which is not observed. We have also discarded this explanation.

\section{Summary}

We have observed multicolor (UBVRI) linear polarization ($P$ and $\theta$) for 35 of stars on the 
direction of NGC 6231.
We have found a very unusual distribution of the angles of the polarization vectors. 
To the North of the cluster the polarization angles are dominated by the direction of the Galactic disk; 
to the South are perpendicular to this orientation.
Close to the core of the cluster, the polarization vectors are well organized in a 
regular pattern, showing a semicircular morphology which  we interpreted 
as a re-orientation of the dust particles by the magnetic field. 
 
We suppose that an energetic event (very probably a supernova) has occurred some time ago and produced a 
shock pattern on the local ISM. 
A supernova is compatible with the evolutionary status of the cluster.
We discuss in  this paper independent confirmations of this event from  studies of the 
interstellar lines \citep{cro01} to the photometric results of the PMS stars given by \cite{sung98}. 

\acknowledgments
We wish to acknowledge the technical support of CASLEO during our observing runs. Also we want to acknowledge the useful discussions with Ana M. Orsatti, Virpi Niemela, Cristina Cappa, Paula Benaglia 
which are greatly appreciated.
We also aknowledge the use of the Torino Photopolarimeter built at Osservatorio Astronomico di Torino (Italy) and operated under agreement between Complejo
Astronomico El Leoncito and Osservatorio Astronomico di Torino. 
We also want to thank to Dr. Michael Corcoran for provide us the Rosat images of the region.

%\documentstyle[12pt,aj_pt4]{article}
%\documentclass{aastex}

%\begin{document}
%\pagestyle{empty}

%\tablewidth{300pt}
%\makeatletter
%\def\jnl@aj{AJ}
%\ifx\revtex@jnl\jnl@aj\let\tablebreak=\nl\fi

%\addtocounter{table}{+3}%
\begin{deluxetable}{lll}
\tablecolumns{3}
%\tablewidth{0pc}
\tablecaption {Polarimetric Observations of stars in NGC 6231}
\tablehead{
  \colhead{Filter} &
  \colhead{ $P_{\lambda} \pm \epsilon_{P}$} &
  \colhead{$\theta_{\lambda} \pm \epsilon_{\theta}$ } \\
     \colhead { } &
     \colhead{$\%$} &
     \colhead{(deg)}
}
\startdata                 
Star   1&  & \\ 
U &  1.24 $\pm$ 0.42 &  58.4 $\pm$  9.3\\
B &  0.84 $\pm$ 0.19 &  47.9 $\pm$  6.5\\
V &  0.66 $\pm$ 0.17 &  25.5 $\pm$  7.1\\
R &  0.94 $\pm$ 0.22 &  26.3 $\pm$  6.6\\
I &  0.99 $\pm$ 0.47 &  32.7 $\pm$ 12.8\\
 Star   6&  & \\ 
B &  0.33 $\pm$ 0.07 &  31.6 $\pm$  6.2\\
V &  0.34 $\pm$ 0.06 &  15.5 $\pm$  5.3\\
R &  0.35 $\pm$ 0.05 &  10.0 $\pm$  3.9\\
I &  0.34 $\pm$ 0.10 &  17.0 $\pm$  8.3\\
 Star   16&  & \\ 
V &  0.36 $\pm$ 0.07 &  20.8 $\pm$  5.7\\
R &  0.37 $\pm$ 0.07 &  15.6 $\pm$  5.0\\
I &  0.31 $\pm$ 0.07 &  24.4 $\pm$  6.3\\
 Star   34&  & \\ 
B &  1.36 $\pm$ 0.12 &  37.4 $\pm$  2.5\\
V &  1.37 $\pm$ 0.14 &  28.4 $\pm$  2.8\\
R &  1.45 $\pm$ 0.17 &  21.7 $\pm$  3.3\\
I &  1.60 $\pm$ 0.24 &  24.8 $\pm$  4.2\\
 Star   70&  & \\ 
U &  1.77 $\pm$ 0.09 & 130.1 $\pm$  1.4\\
B &  0.94 $\pm$ 0.08 & 157.3 $\pm$  2.4\\
V &  1.68 $\pm$ 0.07 & 103.4 $\pm$  1.2\\
R &  1.48 $\pm$ 0.07 & 146.9 $\pm$  1.3\\
I &  1.90 $\pm$ 0.10 & 167.8 $\pm$  1.4\\
 Star   73&  & \\ 
U &  1.82 $\pm$ 0.09 & 140.6 $\pm$  1.4\\
B &  1.59 $\pm$ 0.07 & 164.7 $\pm$  1.2\\
V &  1.08 $\pm$ 0.08 & 101.6 $\pm$  2.1\\
R &  1.87 $\pm$ 0.05 & 154.5 $\pm$  0.8\\
I &  2.53 $\pm$ 0.07 & 169.5 $\pm$  0.8\\
 Star   80&  & \\ 
U &  0.89 $\pm$ 0.15 & 118.9 $\pm$  4.7\\
B &  0.97 $\pm$ 0.09 & 131.2 $\pm$  2.8\\
V &  1.02 $\pm$ 0.10 & 121.9 $\pm$  2.9\\
R &  1.05 $\pm$ 0.09 & 117.1 $\pm$  2.4\\
I &  0.79 $\pm$ 0.16 & 120.3 $\pm$  5.5\\
 Star   102&  & \\ 
U &  0.69 $\pm$ 0.12 &  24.9 $\pm$  5.0\\
B &  0.78 $\pm$ 0.07 &  17.4 $\pm$  2.5\\
V &  0.85 $\pm$ 0.10 &   8.0 $\pm$  3.3\\
R &  0.85 $\pm$ 0.10 &   1.3 $\pm$  3.2\\
I &  0.73 $\pm$ 0.11 &   4.8 $\pm$  4.2\\
 Star   105&  & \\ 
U &  1.06 $\pm$ 0.12 &  93.2 $\pm$  3.4\\
B &  1.19 $\pm$ 0.07 &  88.8 $\pm$  1.7\\
V &  1.09 $\pm$ 0.14 &  83.4 $\pm$  3.6\\
R &  1.03 $\pm$ 0.09 &  88.3 $\pm$  2.4\\
I &  0.99 $\pm$ 0.14 &  83.9 $\pm$  4.1\\
 Star   110&  & \\ 
U &  0.60 $\pm$ 0.14 &  98.7 $\pm$  6.8\\
B &  0.40 $\pm$ 0.13 & 102.4 $\pm$  9.0\\
V &  0.38 $\pm$ 0.12 &  90.6 $\pm$  8.5\\
R &  0.43 $\pm$ 0.10 &  85.1 $\pm$  6.4\\
 Star   112&  & \\ 
U &  0.46 $\pm$ 0.11 & 105.1 $\pm$  6.6\\
B &  0.51 $\pm$ 0.11 & 107.8 $\pm$  5.9\\
V &  0.62 $\pm$ 0.09 & 100.1 $\pm$  4.2\\
R &  0.50 $\pm$ 0.07 &  93.7 $\pm$  4.0\\
I &  0.46 $\pm$ 0.06 &  99.1 $\pm$  3.8\\
 Star   161&  & \\ 
U &  0.29 $\pm$ 0.10 &  25.4 $\pm$  9.4\\
B &  0.38 $\pm$ 0.04 &  41.3 $\pm$  3.1\\
V &  0.46 $\pm$ 0.05 &  43.6 $\pm$  3.4\\
R &  0.31 $\pm$ 0.05 &  40.8 $\pm$  4.9\\
I &  0.13 $\pm$ 0.07 &  38.3 $\pm$ 14.8\\
 Star   166&  & \\
U &  0.69 $\pm$ 0.20 &  43.9 $\pm$  8.3\\
B &  0.69 $\pm$ 0.25 &  49.4 $\pm$  9.8\\
V &  0.69 $\pm$ 0.18 &  44.5 $\pm$  7.5\\
R &  0.54 $\pm$ 0.17 &  40.6 $\pm$  8.9\\
I &  0.54 $\pm$ 0.24 &  36.5 $\pm$ 12.1\\   
 Star   184&  & \\ 
V &  0.55 $\pm$ 0.15 & 177.0 $\pm$  7.6\\
R &  0.44 $\pm$ 0.13 & 170.3 $\pm$  8.0\\
 Star   189&  & \\ 
B &  0.64 $\pm$ 0.20 &  22.5 $\pm$  8.5\\
V &  0.46 $\pm$ 0.11 &  18.4 $\pm$  6.9\\
R &  0.41 $\pm$ 0.09 &  11.1 $\pm$  6.3\\
 Star   194&  & \\ 
U &  0.61 $\pm$ 0.14 &   4.1 $\pm$  6.6\\
B &  0.69 $\pm$ 0.12 &   2.2 $\pm$  4.8\\
V &  0.52 $\pm$ 0.12 & 174.0 $\pm$  6.5\\
R &  0.26 $\pm$ 0.08 & 179.8 $\pm$  8.7\\
 Star   220&  & \\ 
U &  0.63 $\pm$ 0.15 & 116.1 $\pm$  6.7\\
B &  0.62 $\pm$ 0.12 & 130.8 $\pm$  5.4\\
V &  0.63 $\pm$ 0.14 & 112.7 $\pm$  6.4\\
R &  0.65 $\pm$ 0.12 & 105.9 $\pm$  5.1\\
I &  0.50 $\pm$ 0.15 & 109.4 $\pm$  8.3\\
 Star   224&  & \\ 
U &  0.41 $\pm$ 0.14 & 102.1 $\pm$  9.2\\
B &  0.33 $\pm$ 0.08 & 108.6 $\pm$  7.1\\
V &  0.29 $\pm$ 0.07 &  93.8 $\pm$  7.0\\
R &  0.34 $\pm$ 0.07 &  94.3 $\pm$  6.0\\
I &  0.28 $\pm$ 0.10 &  88.2 $\pm$  9.4\\
 Star   232&  & \\ 
U &  0.83 $\pm$ 0.18 & 102.6 $\pm$  6.0\\
B &  0.94 $\pm$ 0.16 & 107.1 $\pm$  4.7\\
V &  0.86 $\pm$ 0.16 & 100.1 $\pm$  5.4\\
R &  0.83 $\pm$ 0.15 & 101.1 $\pm$  5.1\\
I &  0.70 $\pm$ 0.10 &  98.4 $\pm$  4.0\\
 Star   238&  & \\ 
U &  0.39 $\pm$ 0.13 &  95.9 $\pm$  9.1\\
B &  0.36 $\pm$ 0.12 & 111.9 $\pm$  9.4\\
V &  0.36 $\pm$ 0.11 &  94.0 $\pm$  8.7\\
R &  0.41 $\pm$ 0.09 &  89.2 $\pm$  6.4\\
 Star   248&  & \\ 
U &  1.77 $\pm$ 0.51 & 104.2 $\pm$  7.9\\
B &  1.60 $\pm$ 0.17 & 116.5 $\pm$  3.0\\
V &  1.54 $\pm$ 0.19 & 102.7 $\pm$  3.6\\
R &  1.42 $\pm$ 0.25 &  99.9 $\pm$  5.0\\
I &  1.28 $\pm$ 0.43 & 101.5 $\pm$  9.3\\
 Star   253&  & \\ 
U &  0.86 $\pm$ 0.14 & 105.7 $\pm$  4.8\\
B &  1.06 $\pm$ 0.16 & 108.1 $\pm$  4.2\\
V &  0.97 $\pm$ 0.15 &  98.6 $\pm$  4.4\\
R &  0.93 $\pm$ 0.14 &  92.6 $\pm$  4.2\\
I &  0.64 $\pm$ 0.22 &  95.5 $\pm$  9.4\\
 Star   254&  & \\ 
B &  0.96 $\pm$ 0.10 & 112.7 $\pm$  3.0\\  
V &  0.80 $\pm$ 0.14 &  87.8 $\pm$  5.0\\
R &  0.83 $\pm$ 0.12 &  90.5 $\pm$  4.2\\
 Star   259&  & \\ 
U &  0.78 $\pm$ 0.13 & 108.2 $\pm$  4.8\\
B &  1.00 $\pm$ 0.14 & 108.1 $\pm$  3.9\\
V &  0.85 $\pm$ 0.11 & 109.5 $\pm$  3.6\\
R &  0.80 $\pm$ 0.07 & 111.8 $\pm$  2.6\\
I &  0.64 $\pm$ 0.14 & 116.8 $\pm$  6.1\\
 Star   261&  & \\ 
U &  0.58 $\pm$ 0.10 & 108.2 $\pm$  4.9\\
B &  0.57 $\pm$ 0.15 & 105.9 $\pm$  7.2\\
V &  0.52 $\pm$ 0.15 & 101.6 $\pm$  7.8\\
R &  0.51 $\pm$ 0.11 & 102.8 $\pm$  6.0\\
I &  0.43 $\pm$ 0.15 &  99.1 $\pm$  9.4\\
 Star   266&  & \\ 
U &  0.37 $\pm$ 0.07 &  49.5 $\pm$  5.6\\
B &  0.33 $\pm$ 0.06 &  51.8 $\pm$  5.3\\
V &  0.38 $\pm$ 0.08 &  44.7 $\pm$  6.0\\
R &  0.37 $\pm$ 0.05 &  40.8 $\pm$  4.0\\
I &  0.42 $\pm$ 0.09 &  36.2 $\pm$  5.8\\
 Star   272&  & \\ 
U &  0.65 $\pm$ 0.13 & 160.6 $\pm$  5.7\\
B &  0.59 $\pm$ 0.11 & 168.0 $\pm$  5.4\\
V &  0.48 $\pm$ 0.09 & 162.4 $\pm$  5.0\\
R &  0.46 $\pm$ 0.08 & 153.6 $\pm$  5.1\\
 Star   286&  & \\ 
U &  0.64 $\pm$ 0.19 & 152.4 $\pm$  8.2\\
B &  0.52 $\pm$ 0.08 & 142.8 $\pm$  4.3\\
V &  0.35 $\pm$ 0.11 & 132.5 $\pm$  8.4\\
R &  0.39 $\pm$ 0.10 & 128.3 $\pm$  7.1\\
 Star   287&  & \\ 
U &  0.80 $\pm$ 0.12 & 133.5 $\pm$  4.2\\
B &  0.80 $\pm$ 0.10 & 133.6 $\pm$  3.5\\
V &  0.82 $\pm$ 0.09 & 137.1 $\pm$  3.1\\
R &  0.72 $\pm$ 0.11 & 135.5 $\pm$  4.3\\
I &  0.64 $\pm$ 0.10 & 133.4 $\pm$  4.5\\
 Star   289&  & \\ 
U &  0.69 $\pm$ 0.07 & 135.1 $\pm$  2.8\\
B &  0.81 $\pm$ 0.08 & 135.4 $\pm$  2.8\\
V &  0.64 $\pm$ 0.08 & 132.8 $\pm$  3.5\\
R &  0.63 $\pm$ 0.08 & 132.0 $\pm$  3.6\\
I &  0.50 $\pm$ 0.12 & 132.0 $\pm$  6.9\\
 Star   290&  & \\ 
U &  0.65 $\pm$ 0.15 & 172.1 $\pm$  6.4\\
B &  0.64 $\pm$ 0.13 & 169.3 $\pm$  5.9\\
V &  0.65 $\pm$ 0.14 & 157.4 $\pm$  6.0\\
R &  0.58 $\pm$ 0.13 & 157.1 $\pm$  6.5\\
I &  0.51 $\pm$ 0.14 & 163.6 $\pm$  7.7\\
 Star   292&  & \\ 
U &  0.49 $\pm$ 0.11 & 123.1 $\pm$  6.1\\
B &  0.37 $\pm$ 0.09 & 133.4 $\pm$  6.9\\
 Star   HD 152233 &  & \\ 
B &  0.66 $\pm$ 0.12 & 171.7 $\pm$  5.3\\
V &  0.54 $\pm$ 0.10 & 163.3 $\pm$  5.4\\
R &  0.60 $\pm$ 0.11 & 156.6 $\pm$  5.1\\
 Star   HD 152235&  & \\ 
U &  0.97 $\pm$ 0.10 & 105.1 $\pm$  2.8\\
B &  1.00 $\pm$ 0.09 &  99.7 $\pm$  2.5\\
V &  0.97 $\pm$ 0.10 & 101.2 $\pm$  3.1\\
R &  0.90 $\pm$ 0.04 & 100.6 $\pm$  1.4\\
I &  0.71 $\pm$ 0.11 &  95.3 $\pm$  4.3\\
 Star   HD 152248 &  & \\ 
U &  0.55 $\pm$ 0.15 & 120.6 $\pm$  7.6\\
B &  0.49 $\pm$ 0.12 & 125.0 $\pm$  6.6\\
V &  0.59 $\pm$ 0.12 & 119.1 $\pm$  5.9\\
R &  0.49 $\pm$ 0.12 & 111.7 $\pm$  6.9\\
I &  0.52 $\pm$ 0.16 & 105.1 $\pm$  8.3\\
 Star   CPD -47$^{\circ}$ 7733 &  & \\ 
B &  0.31 $\pm$ 0.12 &  96.8 $\pm$ 10.0\\
V &  0.36 $\pm$ 0.10 &  89.5 $\pm$  7.9\\
R &  0.38 $\pm$ 0.09 &  87.1 $\pm$  6.9\\
I &  0.47 $\pm$ 0.09 &  78.1 $\pm$  5.3\\
\enddata
\end{deluxetable}

%\end{document} 

%############################ Segunda tabla ###############################
\addtocounter{table}{+1}
\begin{deluxetable}{c c c c r c}
\tablecolumns{6}
\tablewidth{0pc}
\tablecaption{Comparison between \citet{serk75}\tablenotemark{1}\ \ and this paper}
\tablehead{\colhead{HD} & \colhead{$P_{Serkowski}$} &
\colhead{$\theta_{Serkowski}$} & \colhead{P} &
\colhead{$\theta$} &  ID \\
\colhead{\ \ } & (\%) & degrees & (\%) & degrees & \colhead{Seggewiss } }
\startdata
152248 & 0.66 & 112   & 0.59 $\pm$ 0.12 & 119.1 $\pm$ 5.9 & \\
152234 & 0.70 & 148   & 0.65 $\pm$ 0.14 & 157.4 $\pm$ 6.0 & 290\\
152235 & 0.92 & 110.4 & 0.97 $\pm$ 0.11 & 101.2 $\pm$ 2.6 &  \\
\enddata
\tablenotetext{1}{No error numbers were indicated in the \citet{serk75} paper }
\end{deluxetable}

%\documentstyle[12pt,aj_pt4]{article}
%\documentclass{aastex}
%\begin{document}
\begin{deluxetable}{lcccr}
\tablecolumns{5} 
\tablewidth{0pc} 
%\addtocounter{table}{+2}
\tablecaption {Parameters of the Serkowski fit to the linear polarization data f
or stars in NGC 6231}
\tablehead{
  \colhead{Stellar} &
  \colhead{ $P_{max} \pm \epsilon_{P}$} &
  \colhead{ $\sigma_{1}$ } &
  \colhead{$\lambda_{max} \pm \epsilon_{\lambda_{max}}$ }&
 \colhead{ $\bar{\epsilon}$ } \\
%\colhead{$\sigma_{1}^{*}$} &
%\colhead{$\lambda_{max}^{*} \pm \epsilon_{\lambda_{max}}$ } \\
     \colhead {Identification } &
     \colhead{$\%$} &
     \colhead{  } &
     \colhead{m$\mu$} &
     \colhead{  }
%     \colhead{  } &
%     \colhead{m$\mu$}
}
\startdata             
1     &  0.846 $\pm$   0.112&  1.062&  0.553$\pm$   0.189 & 18.9 \\
6     &  0.355 $\pm$   0.006&  0.169&  0.615$\pm$   0.027 & 10.4 \\
16    &  0.371 $\pm$   0.023&  0.338&  0.542$\pm$   0.065 & 2.3  \\
34    &  1.511 $\pm$   0.077&  0.743&  0.656$\pm$   0.064 & 12.2 \\
70    &  1.616 $\pm$   0.274&  5.933&  0.627$\pm$   0.192 & 367.0\\
73    &  1.953 $\pm$   0.380&  7.997&  0.725$\pm$   0.220 & 199.5\\
80    &  1.036 $\pm$   0.031&  0.577&  0.551$\pm$   0.036 & 8.6  \\
102   &  0.847 $\pm$   0.011&  0.236&  0.581$\pm$   0.015 & 16.2 \\
105   &  1.175 $\pm$   0.027&  0.538&  0.493$\pm$   0.023 & 2.3  \\
110   &  0.476 $\pm$   0.075&  0.932&  0.427$\pm$   0.117 & 6.1  \\
112   &  0.549 $\pm$   0.018&  0.463&  0.536$\pm$   0.031 & 4.1  \\
161   &  0.441 $\pm$   0.141&  1.612&  0.321$\pm$   0.098 & 1.9  \\
166   &  0.699 $\pm$   0.031&  0.209&  0.439$\pm$   0.030 & 2.1  \\        
189   &  0.558 $\pm$   0.131&  0.629&  0.377$\pm$   0.101 & 3.0  \\
194   &  1.018 $\pm$   0.371&  0.987&  0.205$\pm$   0.041 & 2.0  \\
220   &  0.657 $\pm$   0.024&  0.360&  0.506$\pm$   0.035 & 14.4 \\
224   &  0.345 $\pm$   0.027&  0.624&  0.507$\pm$   0.079 & 6.0  \\
232   &  0.911 $\pm$   0.018&  0.220&  0.484$\pm$   0.013 & 2.0  \\
238   &  0.407 $\pm$   0.028&  0.451&  0.549$\pm$   0.083 & 7.9  \\
248   &  1.614 $\pm$   0.043&  0.350&  0.468$\pm$   0.029 & 9.1  \\
253   &  0.985 $\pm$   0.044&  0.524&  0.480$\pm$   0.041 & 7.0  \\
254   &  0.928 $\pm$   0.078&  0.867&  0.460$\pm$   0.091 & 33.5 \\
259   &  0.891 $\pm$   0.035&  0.566&  0.472$\pm$   0.031 & 1.3  \\
261   &  0.588 $\pm$   0.017&  0.257&  0.450$\pm$   0.021 & 1.4  \\
266   &  0.396 $\pm$   0.024&  0.709&  0.589$\pm$   0.069 & 5.6  \\
272   &  0.592 $\pm$   0.063&  0.643&  0.382$\pm$   0.057 &  5.1 \\
286   &  0.533 $\pm$   0.118&  0.864&  0.348$\pm$   0.098 & 8.6  \\
287   &  0.830 $\pm$   0.012&  0.207&  0.474$\pm$   0.011 & 0.6  \\
289   &  0.731 $\pm$   0.032&  0.788&  0.439$\pm$   0.034 &  0.5 \\
290   &  0.666 $\pm$   0.012&  0.163&  0.469$\pm$   0.014 & 6.0  \\
HD 152233  &  0.609 $\pm$   0.066&  0.926&  0.514$\pm$   0.161 & 7.4  \\
HD 152235&  1.020 $\pm$   0.021&  0.386&  0.465$\pm$ 0.015& 1.5  \\
HD 152248  &  0.562 $\pm$   0.029&  0.462&  0.533$\pm$   0.056 & 6.0  \\
CPD -41$^{\circ}$ 7733  &  0.444 $\pm$   0.034&  0.399&  0.852$\pm$   0.095 & 5.6  \\
\enddata
\end{deluxetable}

%\end{document} 

%\section{Figure Captions}

%\begin{figure}
%\caption{3C 244.1, {\em R}-band image with the identifications of the
%structures. The straight line indicates the direction of the
%radio jet.}
%\vspace*{0.25 truein}
%\hspace*{0.25 truein}
%\epsscale{0.7}
%\plotone{fig3.ps}
%\end{figure}
%\end{document} 

%\figcaption[figura1.eps]{Comparison of the common stars observed in two different runs \label{f1}}

%\begin{figure}\caption{Comparison stars in common observed in the two main runs}
%\vspace*{ 0.25 truein}\hspace*{ 0.25 truein}
%ojo el renglo que sigue no estaba en la version que funciona
%\epsscale{0.7}
% \plotone{../figuras/figura1.eps}\end{figure}

%\figcaption[../figuras/figura2.ps]{Projection of the polarization vectors (Johnson V filter) over the sky. Th dot-dashed line is the galactic parallel $b=1^o.1$ \label{f2}}

%\begin{figure}
%\plotone{f1.eps}
%\caption{This is the first figure and it uses sgi9259.eps as
%its EPS figure file. \label{fig1}}
%\end{figure}
 
\clearpage
\begin{figure}\caption{Projection of the polarization vectors (Johnson V filter) over the sky. The dot-dashed line is the Galactic parallel $b=1^o.1$. Star named A is HD 152248, star B is HD 152233 and star C is CPD -41$^{\circ}$7733 \label{fig1}} 
\vspace*{ 0.25 truein}\hspace*{ 0.25 truein}
%\epsscale{0.7}
\plotone{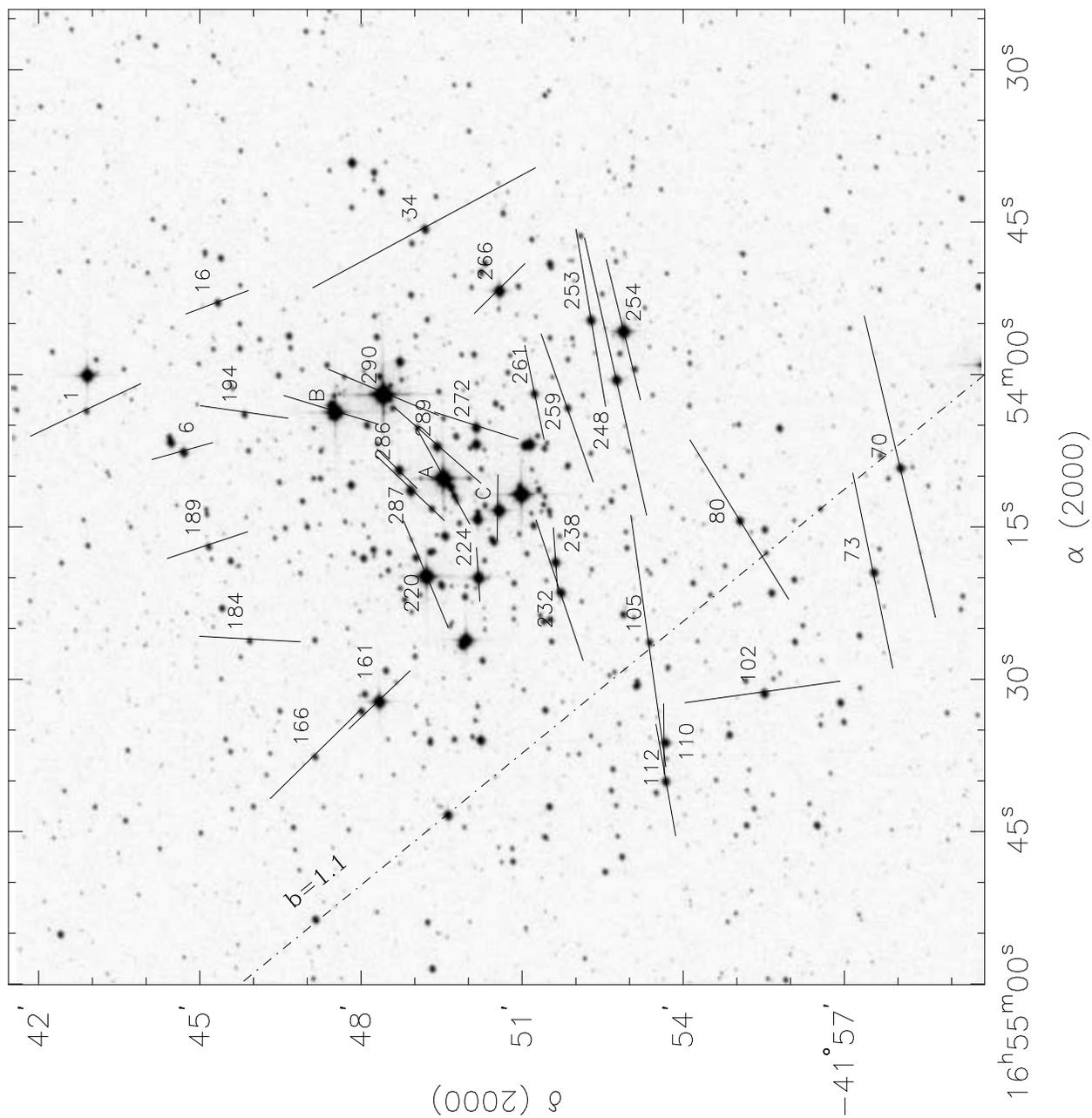}\end{figure}
\clearpage

%\figcaption[feinstein.fig3.ps]{Plot of the observed data for objects showing large departures from the Serkowski law, the solid line is the best fit. \label{f3}}

\begin{figure}
\caption{Plot of the observed data for objects showing large departures from the Serkowski law, the solid line is the best fit.}
\vspace*{ 0.25 truein}
\hspace*{ 0.25 truein}
%\epsscale{0.7} 
\plotone{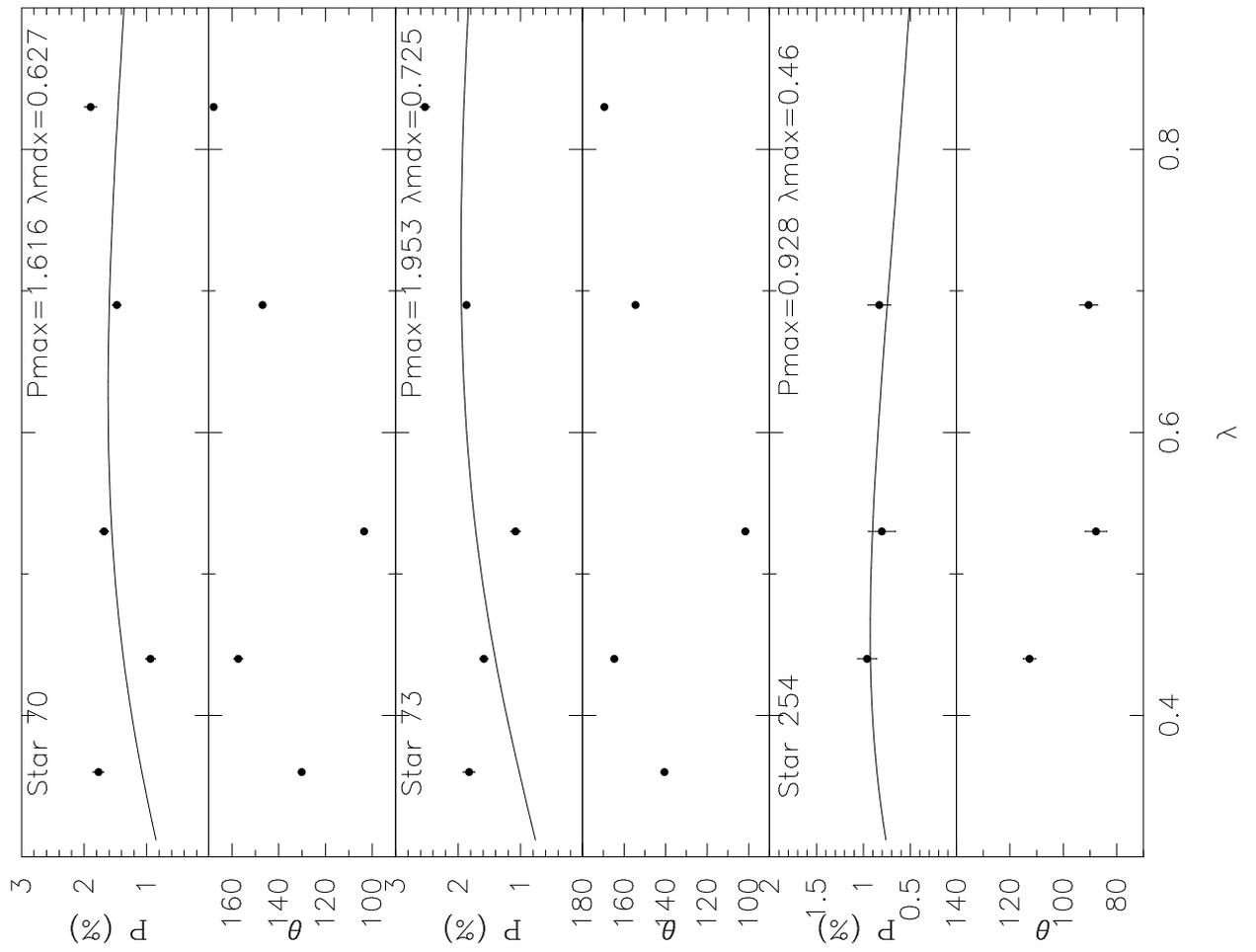}
\end{figure}          

%\figcaption[feinstein.fig4.ps]{ $P_{\lambda max}$ vs $E_{B-V}$ . The solid line is  $P_{\lambda max} = \ 9 E_{B-V}$, and the dashed line is $P_{\lambda max} = \ 5 E_{B-V}$. The upper panel is for stars of group 1, the central panel is for group 2 and the bottom panel is group 3. }

\begin{figure}
\caption{$P_{\lambda max}$ vs $E_{B-V}$ . The solid line is  $P_{\lambda max} = \ 9 E_{B-V}$, the dashed line is  $P_{\lambda max} = \ 5 E_{B-V}$. Top panel is for stars of the first group, middle panel is for the second group and the bottom panel is for the third group}
\vspace*{ 0.25 truein}
\hspace*{ 0.25 truein}
%\epsscale{0.7} 
\plotone{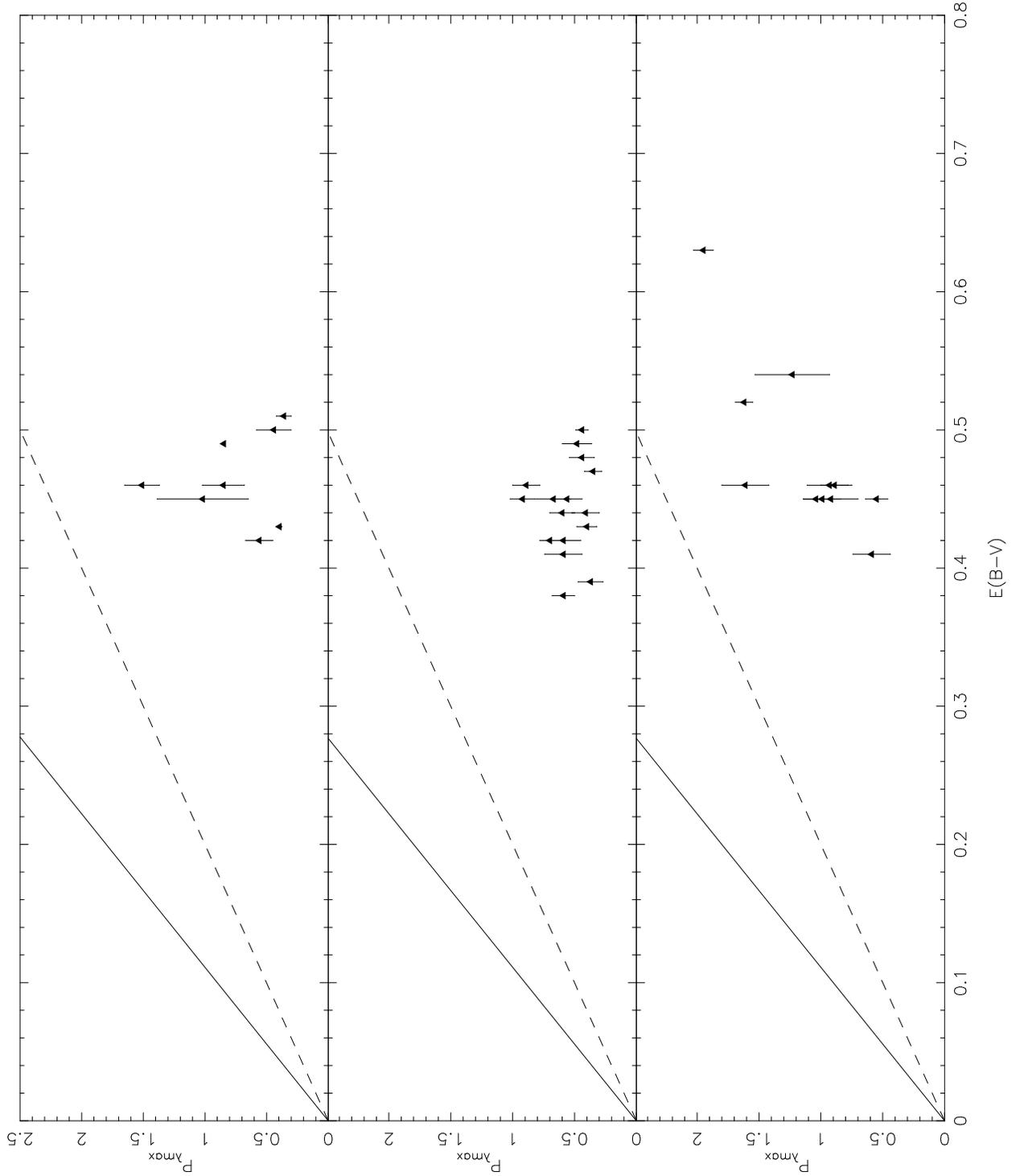}
\end{figure}

%\figcaption[feinstein.fig5.ps]{90 degree plot \label5}

\begin{figure}
\caption{The straight lines from the stars are perpendicular to their polarization vectors. 
These lines seem to converge to a center. Star named A is HD 152248}
\vspace*{ 0.25 truein}
\hspace*{ 0.25 truein}
%\epsscale{0.7} 
\plotone{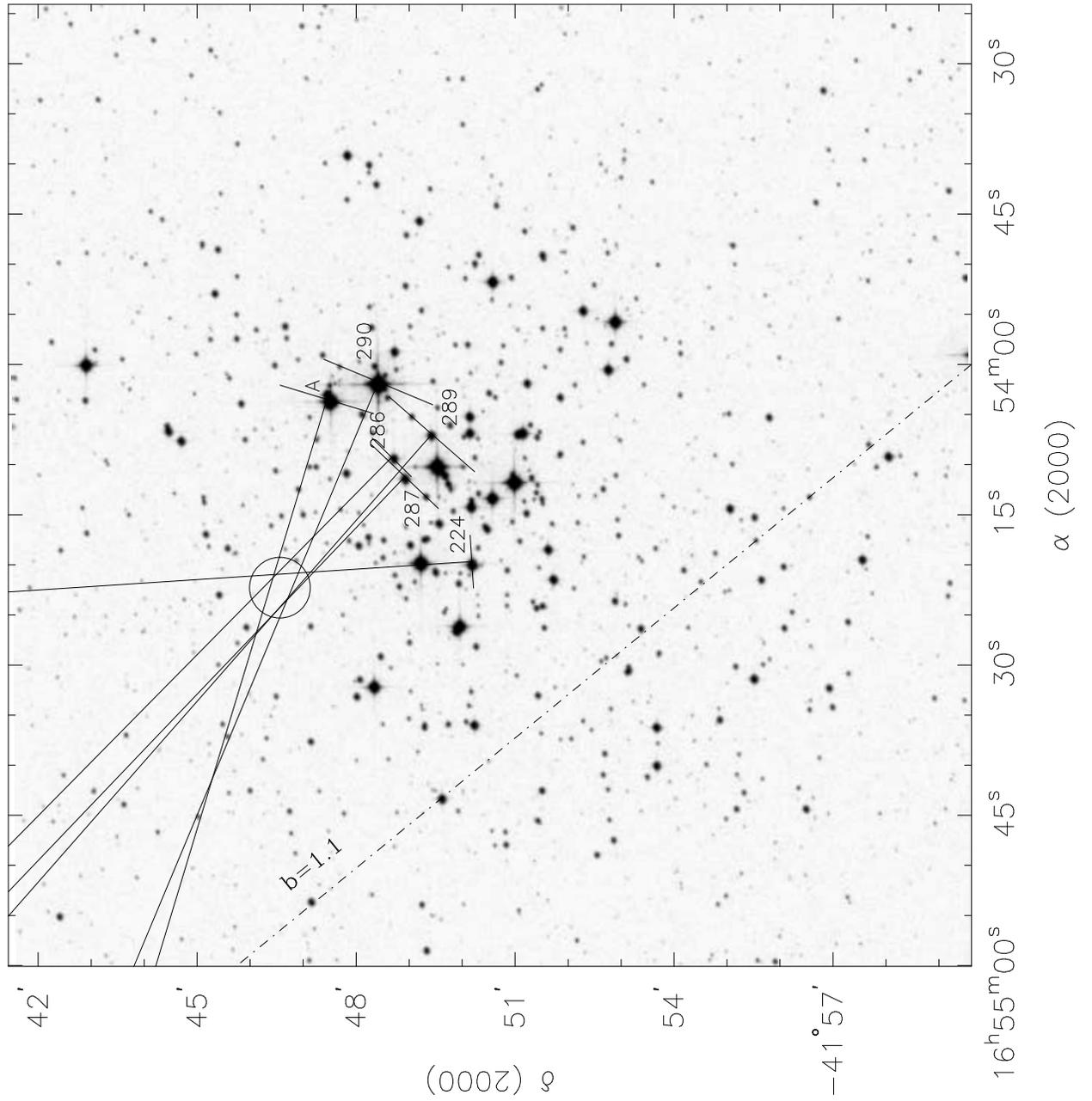}
\end{figure}


\begin{thebibliography}

\bibitem[Balona \& Laney(1995)]{balo95}  Balona, L.A., Laney, 
	C.D. 1995 \mnras, 276,627 
\bibitem[Baume, V\'azquez \& Feinstein(1999)]{bau99}  Baume, G.,
	 V\'azquez, R.A., Feinstein, A. 1999 \aaps, 137,233
%\bibitem[Bok, Bok \& Graham(1966)]{bok66} Bok, B.J., Bok, P.F., 	Graham, J.A. 1966, \mnras, 131,247
\bibitem[Cappa et al.(2002)]{cap02} Cappa, C.E., Goss, W.M., Pineault, S. 2002, \aj, 123,3348
%\bibitem[Corcoran (1999)]{cor99} Corcoran, M.F., Revista Mexicana de Astronom\'{\i}a y Astrof\'{\i}sica,         Workshop on hot stars in open clusters of the Galaxy and the Magellanic Clouds 1999, Vol 8,131
%\bibitem[Corcoran(2003)]{cor03} Corcoran, M.F.,  Private comunication.
\bibitem[Crawford et al.(1989)]{cro89a} Crawford, I., Barlow, M.J., Blades, J.C. 1989 \apj, 336,212
\bibitem[Crawford(1989)]{cro89b} Crawford, I. 1989, \mnras, 241,575
\bibitem[Crawford(1992)]{cro92} Crawford, I. 1992, \mnras, 259,47 
\bibitem[Crawford(2000)]{cro00} Crawford, I. 2000, \mnras, 317,996
\bibitem[Crawford(2001)]{cro01} Crawford, I. 2001 \mnras, 328,1115
%\bibitem[Feinstein \& Ferrer(1968)]{fein68} Feinstein, A., 	Ferrer, O.E. 1968, PASP, 80,410 
\bibitem[Feinstein et al.(2000)]{fein00} Feinstein, C., Baume, G., 
	V\'azquez, R., Virpi, N., Cerruti, M.G. 2000, \aj, 120,1906                                   
\bibitem[Feinstein et al.(2003)]{fein03} Feinstein, C., Baume, G., Vergne, M.,
V\'azquez, R 2003, A\&A, 409, 933
%\bibitem[Garc\'{\i}a \& Mermilliot (2001)]{gar01} Garc\'{\i}a, B., 	Mermilliot, J.C. 2001 \aap, 368,122
\bibitem[Luna(1982)]{lun} Luna, H. 1982, \pasp, 94,695
\bibitem[Manchanda et al.(1996)]{man96} Manchanda, R.K., Polcaro, V.F., 
	Norci, L., Giovannelli, F., Brinkmann, W., Radecke, H.D., Manteiga, M.,
	Persi, P., Rossi, C., 1996, \aap, 305,457
\bibitem[Maronna, Feinstein and Clocchiatti (1992)]{mar92} Maronna, R., Feinstein, C., 
	Clocchiatti, A. 1992, \aap, 260,525
%\bibitem[Mermilliot (1981)]{mer81} Mermilliot, J.C. 1981, \aaps, 	44,467                
\bibitem[Marraco et al. (1993)]{mar93} Marraco, H.G., Vega, E., Vrba, F.J. 1993, \aj, 105,258
\bibitem[Mermilliot \& Maeder (1986)]{mer86} Mermilliot, J.C., 
	Maeder, A. 1986, \aap,158,45
\bibitem[Neckel \& Klare(1980)]{nec80} Neckel, Th., Klare, G. 1980, 
	\aaps, 42,251   
\bibitem[Orsatti et al.(1998)]{orsa98} Orsatti, A.M., Vega, E., 
	Marraco, H.G. 1998, \aj, 116,226   
\bibitem[Perry, Hill \& Christodoulou(1991)]{per91}  Perry, C.L., 
	Hill, G., Christodoulou, D.M. 1991, \aaps, 90,195 
\bibitem[Raboud, Cramer \& Bernasconi(1997)]{rab97}  Raboud, D., 
	Cramer, N., Berneasconi, P.A. 1997, \aap, 325,167 
%\bibitem[Schild, Hiltner \& Sanduleak(1969)]{sch69} Schild, R.E., 	Hiltner, W.A., Sanduleak, N. 1969, \apj, 156,609                                 
\bibitem[Scaltriti et al.(1989)]{scal89} Scaltriti, F., Cellino, A., Anderlucci, E., Corcione, L., Piirola, V. 1989, Societa Astronomica Italiana, Memorie, Vol.60, no. 1-2, p. 243-246.
\bibitem[Seggewiss(1968)]{seg68} Seggewiss, W. 1968, Ver\"off. 
	Astron. Inst. Univ.Bonn, 79
\bibitem[Serkowski et al.(1975)]{serk75} Serkowski, K., Mathewson, 
	D. L., Ford, V. L. 1975, \apj \ 196,261
\bibitem[Shobbrook(1983)]{sho83} Shobbrook, R.R., 1983, \mnras, 	205,1229
\bibitem[St-Louis et al.(1987)]{st87}St-louis, N., Drissen, L., 
	Moffat, A.F.J., Bastien, P., Tapia, S. 1987, \apj, 322,870 

%\bibitem[Sung, Bessell \& Lee(1998)]{sung98} Sung, H., Bessell, M.S., 	Lee, S.W. 1998, \aj, 115,734
\bibitem[Sung et al.(1998)]{sung98} Sung, H., Bessell, M.S., 	Lee, S.W. 1998, \aj, 115,734
%\bibitem[Van Genderen, Bijleveld \& van Groningen(1984)]{van84} 	Van Genderen, A.M., Bijleveld, W., van Groningen, E. 1984, \aaps,58,537                     
\bibitem[Waldhausen et al.(1999)]{wald99} Waldhausen, S., Mart\'{\i}nez R., 
	Feinstein, C. 1999, \aj, 117,2882        
\bibitem[Whittet et al.(1992)]{whittet92} Whittet, D.C.B., Martin, P.G., Hough, 
	J.H., Rouse, M.F., Nailey, J.A.,Axon, D.J., 1992, ApJ, 386,562 

\end{thebibliography}
\end{document}